\documentclass{amsproc}

\usepackage{url}

\usepackage{graphicx}
\usepackage{hyperref}
\usepackage{placeins}
\usepackage{csquotes}

\theoremstyle{definition}

\theoremstyle{remark}

\numberwithin{equation}{section}

\begin{document}

\title{A primer on data-driven modeling of complex social systems}

\author{Alexandria Volkening}
\address{Department of Mathematics, Purdue University, West Lafayette, IN 47907 USA}
\email{avolkening@purdue.edu}
\thanks{In putting together my lecture for the 2021 AMS Short Course, I acknowledge Heather Zinn Brooks, Jonathan Desponds, Simon Freedman, Brian Hsu, Kara Maki, Niall Mangan, and Bridget Torsey for helpful examples in their earlier talks or pointers to references. Special thanks to Jeffrey Humpherys, Rachel Levy, and Thomas Witelski for their minitutorial \cite{LevyMT} on modeling courses at the 2016 SIAM Annual Meeting; I drew from their minitutorial for several of the concepts in Section \ref{sec:perspectives}. Thanks to my Short Course co-organizers Heather, Mason, and Michelle for their encouragement and for being a terrific team, and to Mason for introducing me to the term ``data-driven modeling of complex systems" in the first place.}

\date{\date}


\begin{abstract}
Traffic jams on roadways, echo chambers on social media, crowds of moving pedestrians, and opinion dynamics during elections are all complex social systems. These applications may seem disparate, but some of the questions that they motivate are similar from a mathematical perspective. Across these examples, researchers seek to uncover how individual agents---whether drivers, Twitter accounts, pedestrians, or voters---are interacting. By better understanding these interactions, mathematical modelers can make predictions about the group-level features that will emerge when agents alter their behavior. In this tutorial, which is based on the lecture that I gave at the 2021 American Mathematical Society Short Course, I introduce some of the terms, methods, and choices that arise when building such data-driven models. I discuss the differences between models that are statistical or mathematical, static or dynamic, spatial or non-spatial, discrete or continuous, and phenomenological or mechanistic. For concreteness, I also describe models of two complex systems, election dynamics and pedestrian-crowd movement, in more detail. With a conceptual approach, I broadly highlight some of the challenges that arise when building and calibrating models, choosing complexity, and working with quantitative and qualitative data. 
\end{abstract}

\maketitle

\begin{displayquote}
\emph{A complex system might be defined as a system \\
for which no single model is appropriate.}\\ [5pt]
and
\end{displayquote}

\begin{displayquote}
\emph{As Picasso said of art, a good model \\
``is a lie that helps us see the truth."} \\ [5pt]
(Lee A.\ Segel and Leah Edelstein-Keshet \cite{KeshetSegel})
\end{displayquote}

\section{Introduction}

Traffic jams on roads \cite{Sugiyama2008,Stern2018,Bellomo2011,Jiang2014,Bando1995}, pedestrian crowds \cite{Bailo2018,Helbing1995}, swarming locusts \cite{Ariel2015,Bernoff2020}, animal aggregations  \cite{Dodson2020,Couzin2002,Parrish1999,Lukeman2010keshet,Ballerini2008,Katz2011fish}, collections of cells \cite{Buttenschon2020,VolkeningRev,Giniunaite2020,Glazier1993}, and echo chambers \cite{Sasahara2021,Evans2018,Cinelli2021,Cota2019} are examples of complex systems. In each of these cases, rich, group-level dynamics emerge from the interactions of smaller components---e.g., drivers, people, locusts, animals, or cells---with one another and with their environment \cite{DomenicoWeb,BrockmannWeb}. The interdisciplinary field of \emph{complex systems} \cite{NewmanReview} centers on the questions that arise from these emergent dynamics. Complementing experimental approaches to complex systems, mathematicians develop methods in dynamical systems, topology, network science, numerical analysis, probability, partial differential equations, and many other areas. Here I focus on data-driven mathematical modeling, mainly for complex social systems. My goal for this tutorial chapter is to help provide a starting point for folks who are new to this area, and I reflect on some broad questions and choices that emerge when combining models and data.

\begin{figure}[t]
\includegraphics[width=\textwidth]{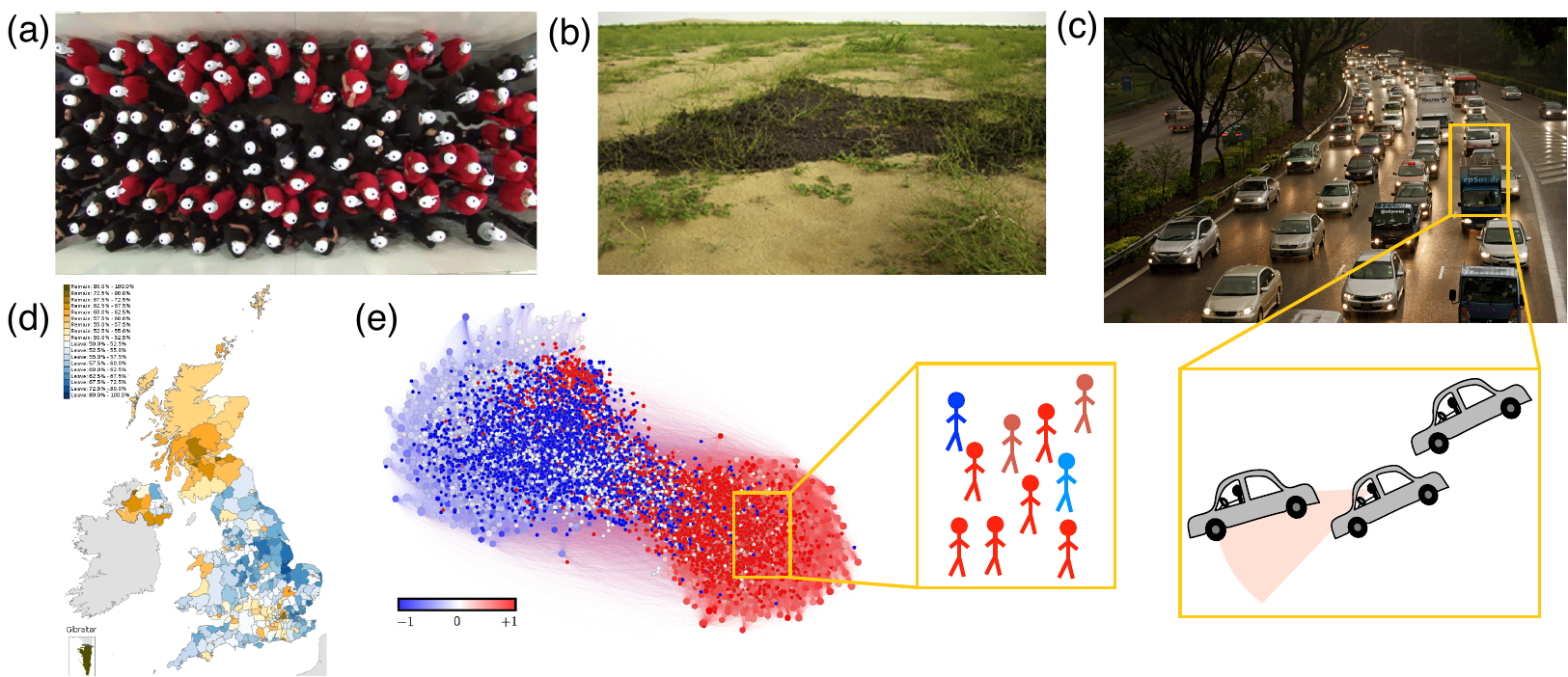}
\caption{Examples of complex social systems. (a) Lanes can emerge from the interactions of pedestrians moving to the right (in black) and left (in red) in a corridor \cite{Sieben2017, Zhang2012}; see Section~\ref{sec:pedestrians}. (b) Locusts form bands as they move over the ground, destroying crops \cite{Cressman2016}. (c) Drivers react to one another and external signals. In an experiment on a circular road \cite{Sugiyama2008}, Sugiyama \emph{et al}. instructed originally equidistant drivers to drive normally. Despite the lack of external signals, a phantom traffic jam formed. This jam or pulse of high density traveled backward relative to the direction that the cars moved; see the supplementary material of \cite{Sugiyama2008} for an animation. (d) Election outcomes \cite{Stolz2016,WikiBrexit} and (e) echo chambers \cite{Cota2019} may emerge from conversations, news coverage, interactions on social media, or other factors. Image (a) adapted (cropped) from \cite{Sieben2017} and licensed under CC-BY 4.0 (\url{https://creativecommons.org/licenses/by/4.0/}); image~(b) reproduced from \cite{Cressman2016} with permission from Elsevier, Copyright (2016) Elsevier Inc.; image (c) reproduced from \cite{WikiTraffic} and licensed under CC-BY 2.0 (\url{https://creativecommons.org/licenses/by/2.0/}); image (d) reproduced from \cite{WikiBrexit} and licensed under CC BY-SA 3.0 (\url{https://creativecommons.org/licenses/by-sa/3.0/deed.en}); image~(e) reproduced from \cite{Cota2019} and licensed under CC-BY 4.0 (\url{http://creativecommons.org/licenses/by/4.0/}); I added the boxes and cartoons with detail in (c) and (e). \label{fig:examples}}
\end{figure}

Figure~\ref{fig:examples} highlights several complex social systems, ranging from traffic flow \cite{Sugiyama2008,Bellomo2011,Nagel1992} to Brexit voter dynamics \cite{Stolz2016}; I also recommend the supplementary material of \cite{Sieben2017,Sugiyama2008} and the websites \cite{LocustVideo,EchoDemo} for related animations. Across these applications, one interesting feature is the common challenges that they raise from a modeling perspective. For example, in each of the images in Figure~\ref{fig:examples}, a researcher may want to characterize alignment. This can be physical alignment, with pedestrians, locusts, or drivers adjusting how they move in response to other individuals or obstacles in their environment. A different type of alignment is present in Figure~\ref{fig:examples}(d)--(e): people are forming opinions and may be influenced to align with (or against) the beliefs of others. Another thread in complex systems is heterogeneity \cite{Miller2007}: each person, animal, or social-media account in Figure~\ref{fig:examples} is unique. Guided by the data available, each modeler must choose how much detail to include. Should we model voters as having a binary opinion (e.g., ``for Brexit" or ``against Brexit") or allow opinions to live on a spectrum? Changes in behavior are also present: for example, in evacuation conditions, an emotional contagion can propagate through a crowd, changing how pedestrians act \cite{Bosse2009,Bosse2011,Tsai2011,bertozzi2015contagion}.

Higher-order interactions are widespread in complex systems: peer influence and social reinforcement from multiple friends may cause someone to change their opinion or adopt a new technology, when an isolated or pairwise interaction might not \cite{Oster2012,Bandiera2006,Lacopini2019,Guilbeault2018,Schelling1973}. In a related vein, the presence of short- and long-range interactions in complex systems leads to rich dynamics. In Figure~\ref{fig:examples}(c), drivers are interacting locally, basing their acceleration on the cars near them. The addition of autonomous vehicles allows for long-range dynamics. Stern \emph{et al.} \cite{Stern2018} have shown that judiciously modulating the speed of one autonomous vehicle can result in the disruption of phantom traffic jams and improved fuel usage in some experiments. (Phantom traffic jams are jams that appear to emerge from drivers, rather than through external forces \cite{Jiang2014,Stern2018}.)

Modeling complex social systems stems from and leads to questions that are of societal and mathematical interest. From an applied perspective, in the case of traffic flow, we might want to shed light on what driver behaviors cause jams or suggest how to use external controls---e.g., time-dependent gating at ramps---to improve traffic. Models can also provide insight into how echo chambers form or suggest interventions to help dissipate divisions. These goals fall into the framework of seeking to understand normal and altered agent interactions, and to predict resulting group-level features. From a mathematical perspective, modeling complex systems can be a starting point to drive the development of new methods and inspire researchers to combine subfields in novel ways.

Motivated by the breadth of complex social systems, my tutorial lecture ``Data-driven modeling" kicked off the 2021 American Mathematical Society (AMS) Short Course on Mathematical and Computational Methods for Complex Social Systems, and this chapter is an offshoot of that talk. Many of the figures in this tutorial are related to slides in my presentation; these slides and my talk recording are available at \cite{VolkeningVideo}. Following the structure of my Short Course presentation, this chapter has three main parts and takes a conceptual approach throughout. First, in Section~\ref{sec:resources}, I highlight some resources, including those that I drew on when preparing my lecture. Second, in~Sections \ref{sec:perspectives}--\ref{sec:modeling}, I overview mathematical modeling and discuss some of the approaches, challenges, and choices that can arise when working with data. Third, in Section \ref{sec:case}, I discuss two case studies---election forecasting and pedestrian movement---in more depth.

Mathematical modeling is a big field, and data-driven modeling can be defined in different ways. The array of approaches that modelers can choose from is a strength, since different perspectives contribute in complementary ways to our understanding of complex systems. As a central theme, I want to acknowledge these choices and use the quotations from Segel and Edelstein-Keshet \cite{KeshetSegel} at the start of this chapter as a guide. The abundance of modeling approaches to complex systems, coupled with their multidisciplinary nature, also means that communication is more challenging; researchers may not mean the same thing when they say the same term. With this in mind, I discuss some of the things that I---from my perspective as an applied mathematician and math biologist---consider when I think about modeling complex systems. There are many, many perspectives on modeling, and this tutorial represents one, informed by the references herein.

\section{Some Resources on Modeling}\label{sec:resources}

I point out some resources below, including the materials that I drew on for my Short Course lecture \cite{VolkeningVideo}.

\subsection{Free Online Resources}

The websites \cite{BrockmannWeb,DomenicoWeb} provide dynamic examples of research in complex systems and are an excellent place to gain intuition and explore this field. The Society for Industrial and Applied Mathematics (SIAM) hosts two modeling handbooks \cite{SIAMmodelingComp,SIAMmodeling}; and SIAM and the Consortium for Mathematics and its Applications provide guidelines on teaching mathematical modeling \cite{GAIMME}. Humpherys, Levy, and Witelski organized a very useful minitutorial discussing graduate and undergraduate education in modeling at the 2016 SIAM Annual Meeting; both their slides and a recording of their presentation are available online \cite{LevyMT}. Kutz and Brunton have posted a rich collection of videos \cite{BruntonYouTube,KutzYouTube} on YouTube, discussing topics including data-driven model discovery. For a demonstration of how to go from a biological paper to making simplifications to building different models, my tutorial lecture \cite{VolkeningBioVideo} for a broad audience may be of interest. Also geared toward a biological audience, the course ``What do Your Data Say?" \cite{What} includes a large collection of video lectures with a statistical, data-driven perspective. To see examples of research talks related to modeling complex systems, I highlight some of the BIRS workshop videos \cite{BIRS} (this collection from the University of British Columbia library contains a wider selection of topics than just modeling), as well as videos in the virtual SIAM Data Science minisymposium ``Topological Techniques and Data-Driven Modeling in Complex Systems" organized by Brooks and Porter \cite{SessionYouTube}.

\subsection{Books}
I found the books \cite{Kutz2016,Kutz2013,brunton2019book} to be especially helpful as I developed my Short Course lecture, and the book \cite{KeshetSegel} by Segel and Edelstein-Keshet provides the quotations that open this chapter. Additional books related to complex systems and modeling include \cite{Mitchellbook, Thurner,BoccaraBook,Miller2007}.

\subsection{Publicly Available Data}
Accessing data can be a challenge in complex-systems research. As a starting point, I highlight some publicly available data for a few specific applications here. For studies on elections and political opinions in the United States, I recommend the breadth of polling data aggregated by FiveThirtyEight \cite{LatestPolls}. HuffPost Pollster also curates a broad collection of public polls, with a search bar for finding data \cite{HuffPostPollster,HuffPostAPI}. At a finer scale, the 2016 presidential election results in California are available at the precinct level from the \emph{Los Angeles Times} \cite{LATimesData}. Ciocanel, Topaz, and other researchers through the Institute for the Quantitative Study of Inclusion, Diversity, and Equity (QSIDE) \cite{QSIDE} developed a large-scale database (called JUSTFAIR, for Judicial System Transparency through Federal Archive Inferred Records) holding over $500,000$ federal district court records \cite{Justfair}. Data from the social-media platform Twitter, as well as tutorials, are available from sources including \cite{Storywrangler,Storyweb,TwitterKalt,Twitter}.

\section{Some Perspectives on Data and Models}\label{sec:perspectives}

Because terminology can vary across fields, I survey some terms for describing models (Section~\ref{sec:models}) and data (Section~\ref{sec:data}), and then define data-driven modeling for the purposes of this chapter (Section~\ref{sec:dataModels}). If you are coming to this tutorial with an applied question that you want to address, I encourage you to keep your complex system in mind as you read---what are the parameters in your system, what data could you use to constrain your model, and at what scale do you want to make predictions or describe the system? If you are a mathematician new to modeling, what mathematical challenges does thinking from the perspective of complex systems raise? If you are from a different disciplinary background than mine, how does what we mean by ``data-driven modeling" differ from and complement each other? And, if you happen to be a modeler who---like me---was introduced to modeling through research, it might be interesting to reflect on how we teach modeling.

\subsection{Types of Models}\label{sec:models}

\begin{figure}[t]
\includegraphics[width=\textwidth]{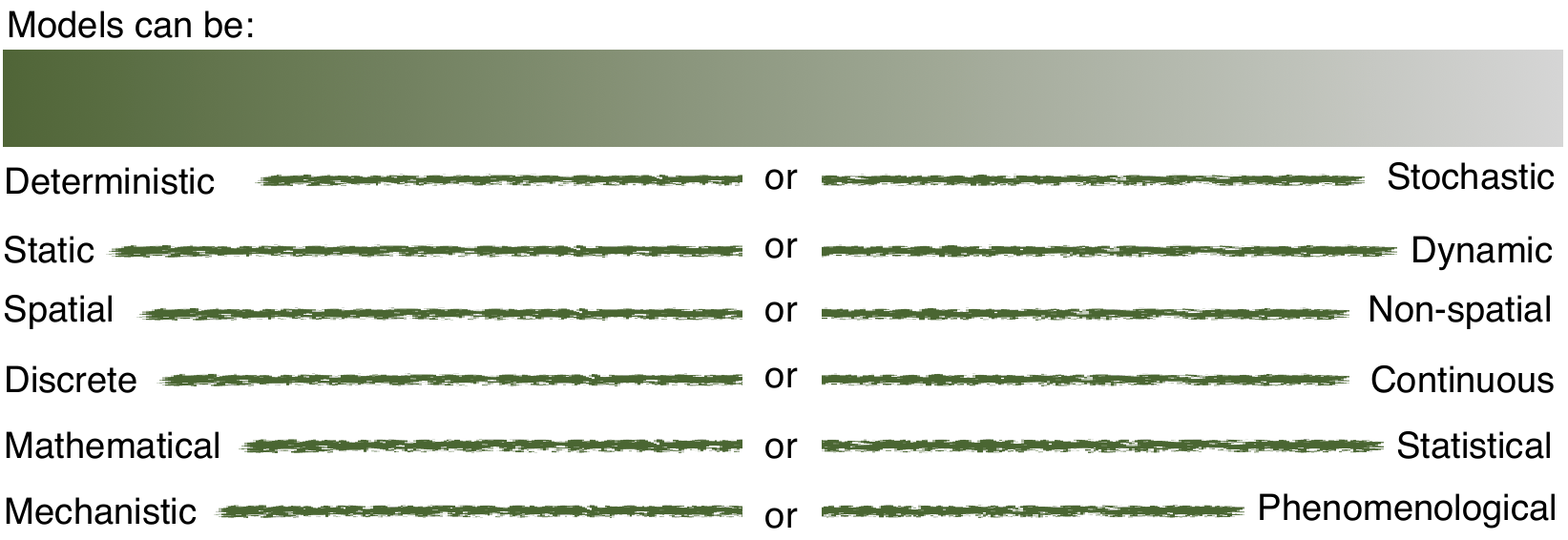}
\caption{Example modeling perspectives \cite{VolkeningVideo}. Many models fall somewhere in the middle of each of these scales. For example, a model may have both stochastic components (e.g., stochastic rules for when new pedestrians enter the corridor in Figure~\ref{fig:examples}(a)) and deterministic components (e.g., differential equations for pedestrian movement). Models can be discrete or continuous in many ways: they can be discrete in terms of types of opinions (e.g., Republican or Democratic voting opinions in the United States), physical space, or time, for example. The distinction between phenomenological and mechanistic models is difficult, and folks have different opinions on what this means, as I discuss in Section~\ref{sec:models}.\label{fig:perspectives}}  
\end{figure}

The term ``model" means different things in different fields. In the life sciences, ``model" may refer to a model organism (e.g., zebrafish, fruit flies, or worms) \cite{LevyMT} or a schematic hypothesizing the relationship between things. In mathematics, we may think of models that take the form of differential equations or stochastic rules, for example. Mathematical models are described using many terms, and I include a few in Figure \ref{fig:perspectives}. Figure~\ref{fig:perspectives} also highlights some of the initial choices that modelers face, often constrained by their data. Importantly, the distinctions in Figure~\ref{fig:perspectives} are not sharp: models often fall on a spectrum and this can depend heavily on the perspective that one takes.

Models can be described as deterministic or stochastic; stochastic models include variability. Depending on their goals and data, researchers must choose whether to build models that are static (time-independent) or dynamic. Similarly, scientists are faced with the choice of building models that are spatial or non-spatial. Do we need to understand where individual cars are located on a road, or is it sufficient to know how the number of cars evolves? Multiscale approaches are also possible, and I provide an example for the case of pedestrian movement in Section~\ref{sec:pedestrians}. We can think of models as being discrete or continuous in time or in space (e.g., so-called ``on-lattice" or ``off-lattice" microscopic models; see Section~\ref{sec:pedestrians}), but models can also be discrete in terms of types of agents; for example, do we assume voters live on an ideological spectrum or assign them a binary opinion? Whereas the choice of making a model discrete or continuous in space and time is often a choice of mathematical and computational implementation, the choice of modeling agents as having discrete or continuous features can be particularly meaningful from the perspective of the application. Understanding how choices of implementation impact model predictions is an important area of research (e.g., \cite{Kursawe2017}), as is uncovering how different modeling approaches---such as microscopic and macroscopic (see Section~\ref{sec:pedestrians})---are related (e.g., \cite{Bernoff2011,Carrillo2021,Bodnar}).

Some researchers distinguish between mathematical and statistical models, and others see statistical models as a type of mathematical model. A related categorization is phenomenological or mechanistic. These are difficult distinctions, and, in my opinion, scientists use the terms ``phenomenological" and ``mechanistic" in different ways. Mechanistic models of complex systems get at the mechanism underlying agent behavior. For example, the drivers in Figure~\ref{fig:examples}(c) want to avoid running into one another, and we could model this by specifying repulsive forces between cars. This model can be seen as phenomenological since it describes the affect (e.g., drivers avoid one another) without getting at the mechanism of how the repulsion occurs. If we modeled the physics of the vehicles, the vision cone of individual drivers, and each driver's internal decision process, this would be more mechanistic. However, what are the variables in a model of how people make decisions? This is in some sense a phenomenological model as well, raising further questions that involve neuroscience. I suggest that the meanings of ``mechanistic" and ``phenomenological" depend on the question that we want to answer and the perspective from which we are studying an application. In my opinion, many models are mechanistic at one scale, and phenomenological as soon as we step deeper into the complex system.

 \subsection{Types of Data}\label{sec:data}
 
 The methods that modelers use to build predictive models that balance model and data complexity look different depending on the form of their data. However, the core concepts are the same when building and validating data-driven models if we look more closely, and, for this reason, I overview some types of data here. For example, data may be quantitative (e.g., the speed of the $i$th car in Figure~\ref{fig:examples}(c)) or qualitative (e.g., the presence of lanes emerging from pedestrian behavior in Figure~\ref{fig:examples}(a)). See Section~\ref{sec:elections} for an example of modeling with quantitative data, and Section~\ref{sec:pedestrians} for a discussion of the challenges that qualitative data introduce to the modeling process. Textual data also emerge from many complex social systems (e.g., \cite{Storywrangler,Minot2022}).

Sometimes we find ourselves with so much data that we cannot open the files, and other times there is nearly no data. In the first case, the ``black-box" modeling approaches that I discuss in Section~\ref{sec:dataModels} may be useful; for example, if we are working with a huge set of tweets, we could complete some data analysis to identify meaningful categories of accounts. If we are working with large sets of qualitative data (e.g., many images), this may motivate the development of new computational and mathematical approaches for extracting quantitative information from our data. 
On the other hand, if we have nearly no data, it can be challenging to know where to start. In this case, it is a matter of making many simplifications (and being actively aware of the choices that we make in this process), so that the number of assumptions that we build into our model is balanced with the small amount of data available.

On a related note to amount, data for some complex systems describe rare events. For example, a model may be fit to measurements of average traffic flow, but how do we account for events that are relative outliers, like car accidents? In the case of election forecasting, we might judge a model as wrong if, despite giving Candidate $A$ a $75$\% chance of winning and Candidate $B$ a $25$\% chance of winning, Candidate $B$ wins. The reality is that we do not have enough information to determine whether the model is good or bad. Forecasts are more meaningfully judged in aggregate across many elections, but limited polling data are available. Like models, data can also be time-independent (e.g., a Twitter followership network at one snapshot in time) or dynamic (e.g., the timeline of tweets from a given account) and spatial or non-spatial. The initial form of data is often messy, and in some cases a large portion of the time that researchers spend modeling complex systems is focused on cleaning \cite{SIAMmodelingComp}, gathering, and tracking down the oddities in their data.

All forms of data can have bias and require human choices, particularly in the case of complex social systems. I point the reader to the chapter \cite{PorterData} by Porter and references therein in this volume for a discussion of data ethics. Importantly, just because data exist does not mean they should be used, and as the author mentions in \cite{PorterData}, determining when to use or not use data is a critical step in research on complex social systems. Modelers need to be actively aware of the choices that they make when handling data, and of the presence of any choices made prior to the time that they gained access to the data. For example, if we are interested in understanding the online conversation about a recent event, we might start by downloading a large set of tweets using hashtags associated with that event. There are multiple choices wrapped up in this process, and I name a few here \cite{Cihon2016,Morstatter2013,Tufekci2014,Tien2020}. First, we chose one of many social-media platforms, so our analysis will be specific to the groups that use Twitter \cite{Tufekci2014}. Second, we had to select what hashtags to search for and how we would identify tweets ``associated with" our recent event \cite{Morstatter2013,Tien2020}. Third, while the Twitter API provides a rich sample of tweets, it is not fully clear how this selection is made \cite{Morstatter2013}. All of these choices will affect the results of our model.

\subsection{Perspectives on Modeling with Data}\label{sec:dataModels}

\begin{figure}[t]
\includegraphics[width=\textwidth]{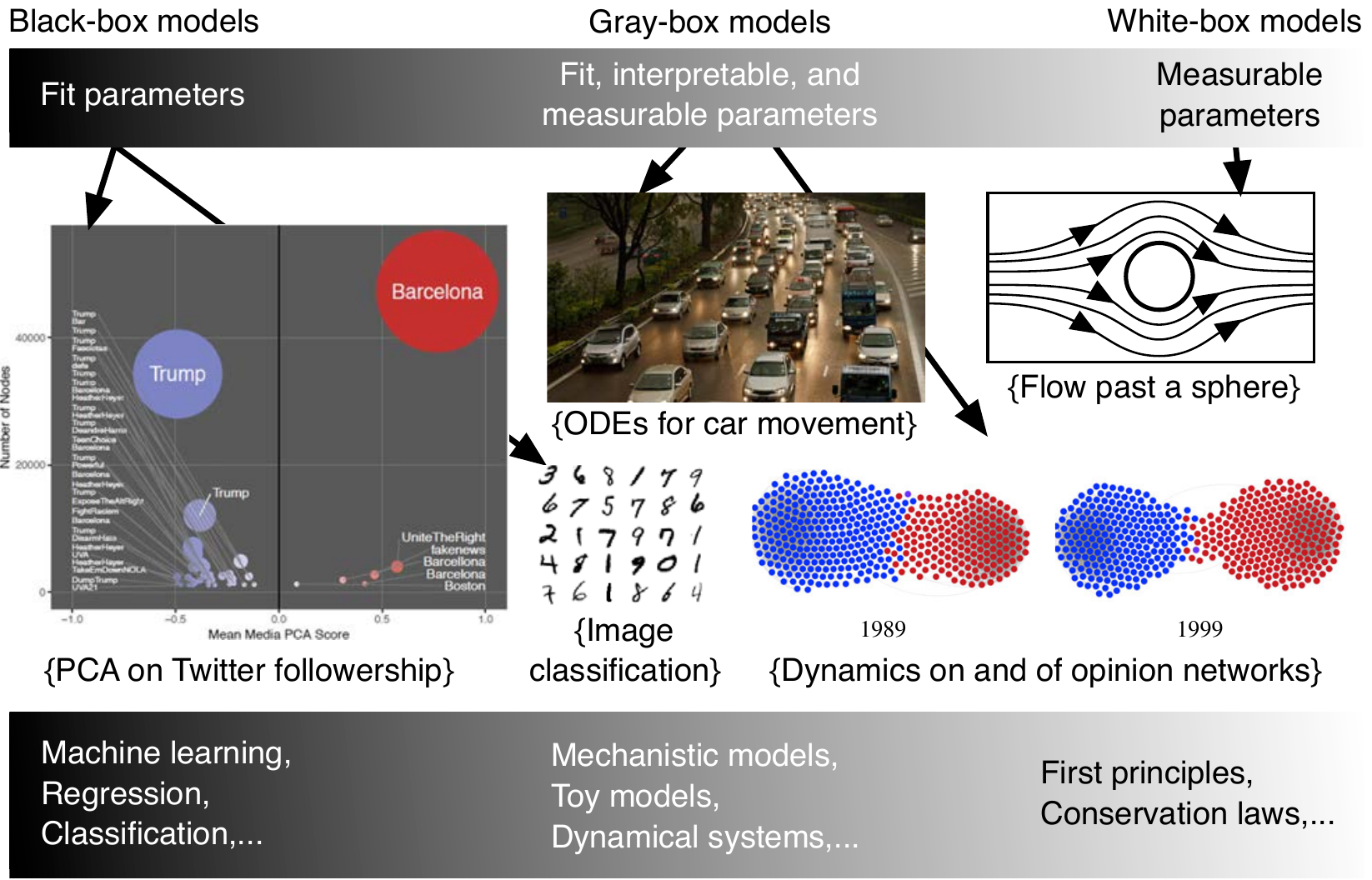}
\caption{Shades of modeling with data \cite{LevyMT}. Black-box, gray-box, and white-box models depend on data to varying degrees and have different relationships with parameters. Black-box modeling approaches rely on data and often have internal parameters, while white-box models are largely dictated by first principles and have measurable parameters (e.g., conductivity of a material). Gray-box modeling involves visible, interpretable parameters that are fit, specified, or measured using data. All of these approaches require domain expertise. As some examples, I highlight principal component analysis (PCA) applied to media followership on Twitter \cite{Tien2020}, and recognizing handwritten numbers \cite{Lecun1998} (black-box modeling); deterministic models of traffic flow and game-theoretic models of opinion dynamics on networks \cite{Evans2018} (gray-box modeling); and fluid flow past a sphere (white-box modeling). PCA--Twitter image and opinion-network images reproduced from \cite{Tien2020} and \cite{Evans2018}, respectively, and licensed under CC-BY 4.0 (\url{https://creativecommons.org/licenses/by/4.0/}); image-classification image reproduced from \cite{Lecun1998} with permission, Copyright (1998) IEEE; traffic image reproduced from \cite{WikiTraffic} and licensed under CC-BY 2.0 (\url{https://creativecommons.org/licenses/by/2.0/}). \label{fig:shades}}
\end{figure}

In their 2016 SIAM Annual Meeting minitutorial \cite{LevyMT}, Humpherys, Levy, and Witelski discussed a useful classification of models based on ``shades of model uncertainty". As I highlight in Figure~\ref{fig:shades}, black-box, gray-box, and white-box models have different levels of dependency on data \cite{LevyMT}, and their parameters mean different things. According to the classification system in \cite{LevyMT}, black-box models are based heavily on data and can be thought of as maps between inputs and outputs; these models include regression, classification, and machine learning. For example, Tien \emph{et al.} applied \cite{Tien2020} principal component analysis to Twitter data (the input) to distinguish groups of accounts (the output) based on their media followership. The parameters in black-box models may be internal or hidden, and it is the model output---rather than the model structure---that is often of most interest. On the other hand, white-box models are based on first principles; these include equations from physics, such as those describing fluid dynamics or optics \cite{LevyMT}. The parameters in white-box models are measurable, and examples are viscosity and conductivity. Gray-box models depend on a combination of data, first principles, and domain expertise. For example, an ordinary differential equation (ODE) model for driver movement could include equations for velocity and acceleration that are based on phenomenological descriptions of repulsion and attraction between cars (i.e., domain expertise) and measurements of speeds (i.e., data).

The distinctions in Figure~\ref{fig:shades}, like the distinctions in Figure~\ref{fig:perspectives}, are not perfect. For example, equation-learning and model-selection approaches (e.g., \cite{Mangan2017,Brunton2016,Nardini2021,Kemeth2022,Mangan2016}) might be thought of as ``dark gray". It is also important to keep in mind that there are choices present and domain expertise needed across the spectrum in Figure~\ref{fig:shades}. This is especially true when working with data from complex social systems, since even the data that are selected for training black-box models rely on a modeler's choice to use those data \cite{PorterData}. For the purposes of this tutorial, I thus think of data-driven modeling as being mathematical modeling that is driven by data, motivated by a given question, and combined with domain expertise. This encompasses developing predictive, mechanistic models based on data; equation learning and model selection\footnote{Equation learning and model selection---sometimes referred to as ``data-driven modeling"---are outside the scope of this survey. See, for example, \cite{Mangan2017,Brunton2016,Nardini2021,Kemeth2022,Mangan2016} for more discussion of these topics.}; machine learning, regression, or classification to understand data; and using models to raise questions and drive further data collection. Both black-box and gray-box models fit this description, but I predominantly focus on gray-box models in this survey, though again I stress that the distinctions are not sharp.

\section{Challenges, Choices, and Creativity in Data-Driven Modeling} \label{sec:modeling}

Data-driven modeling involves creativity and choices, informed by the modeler's driving question, data, and domain expertise. 
In Section \ref{sec:process}, I provide an example modeling process and highlight some of the places where modelers make choices. In Section \ref{sec:complexity}, I then discuss challenges related to data and model calibration. See Sections~\ref{sec:elections} and \ref{sec:pedestrians} for illustrations of these topics for two specific applications. I take a conceptual approach throughout.

\subsection{Building Data-Driven Mathematical Models}\label{sec:process}

As an example data-driven, gray-box modeling process, we might follow the steps below \cite{GAIMME,SIAMmodeling,SIAMmodelingComp}:
\begin{enumerate}
\item formulate our broad motivation and specific goals
\begin{itemize}
\item get to know the application area or talk to domain experts
\item search for data (qualitative or quantitative) and prior work
\item identify hypotheses to be tested or proposed and questions to be ``answered" or raised
\end{itemize}
\item come up with a plan for building and evaluating our model
\begin{itemize}
\item determine baseline assumptions and simplify where possible
\item identify our variables, parameter names, timescales, and units
\item specify the values of measurable parameters and determine what parameters need to be fit
\item handle formatting, cleaning, and quantifying our data as needed
\item break our data into sets for fitting parameters, testing, and predicting
\end{itemize}
\item simulate, analyze, and use our model
\begin{itemize}
\item identify remaining parameter values using data for fitting
\item validate our model on test data
\item perform a sensitivity analysis or bifurcation analysis if possible
\item use our model to gain intuition, raise questions, and make predictions
\item communicate results to an interdisciplinary audience
\item iterate to improve
\end{itemize}
\end{enumerate}
These steps are not necessarily linear and data-driven modeling is iterative \cite{GAIMME,SIAMmodeling,SIAMmodelingComp}. The starting point may be data, domain expertise, or questions, and Step~($1$) involves research to begin filling in gaps in our knowledge of these three areas and to formalize our goals. I often review literature in Step ($1$) with Step ($2$) in mind, tagging papers with quantitative data that I can use later for parameter fitting and noting studies that show alternative experimental conditions that could be used for model testing. Steps ($2$) and ($3$) then treat complementary parts of model building. 

In Step~($2$), we select our overall approach and the variables, parameters, group dynamics, and agent behaviors in which we are most interested. This means making choices related to the concepts in Figures~\ref{fig:perspectives} and \ref{fig:shades}: for example, if we are studying traffic flow on a stretch of roadway, will we track the number $N(t)$ of cars on the road in time, or the position $\textbf{x}_i(t)$ and velocity $\textbf{v}_i(t)$ of each vehicle $i$ in time? If we are accounting for driver differences, will we assume that each driver's phenomenological ``level of cautiousness" is time-dependent or static? It is important to make these choices in a way that accounts for the complexity of the problem and our data, so Step ($2$) involves making a plan for how we will use data to develop (or train, or fit) our model and later test (or validate) our model, as I discuss in Section~\ref{sec:complexity}. At the end of Step ($2$), our model is written down (e.g., as a system of differential equations on paper or as a set of stochastic rules in code).

In Step ($3$), we turn to filling in parameter names with parameter values, setting initial conditions, and determining our boundary conditions, as needed. Step~($3$) involves validating our model to test its predictive value and performing various analyses to check how sensitive our model is to uncertainty in parameter values, initial conditions, boundary conditions, or data. Depending on the form of our model, we may be able to perform a bifurcation analysis to understand how changes in parameters influence our results. We may also brainstorm alternative ways of judging model output and comparing this with data, since how we choose to describe model output can impact how we interpret our results. At the end of Step~($3$), we use our model to gain understanding and, if possible, suggest new experiments, resulting in model-driven data collection.

More broadly, Step~($1$) is where we realize that a model can help us accomplish our goals, Step~($2$) is the place where we build the model structure, and Step~($3$) is where we test and prod this structure. Data enter the picture in Step~($1$) as motivation. In Steps~($2$) and ($3$), we work closely with data to build, test, and use our model in a way that balances model and data complexity to accomplish our goals. In the remainder of this tutorial, I focus primarily on the later parts of Step ($2$) and broadly discuss the early parts of Step ($3$). To learn more about some of the analyses and computational approaches possible in Step ($3$), I suggest the books \cite{SmithBook,StrogatzBook,Kutz2013,KeshetSegel}.

\subsection{Balancing Model and Data Complexity}\label{sec:complexity}

While data-driven models take many forms and scientists use a range of methods to understand them, the overarching theme of balancing model and data complexity is present throughout. Depending on our goals and data, what modeling approach do we choose? How do we build a data-driven, \emph{data-appropriate} model? If we have access to a wealth of domain expertise and a rich set of data, it may make sense to build a complex model, since, in this case, the majority of the model will be purely descriptive, framing known agent interactions in a mathematical way. The new hypothesis that we are testing, along with its few parameters, enters the picture as our assumption. On the other hand, if we are leading the way to model a poorly understood complex system, our model needs to be very simple, again so that the assumptions and hypotheses that we introduce match the amount of data available.

\begin{figure}[t]
\includegraphics[width=\textwidth]{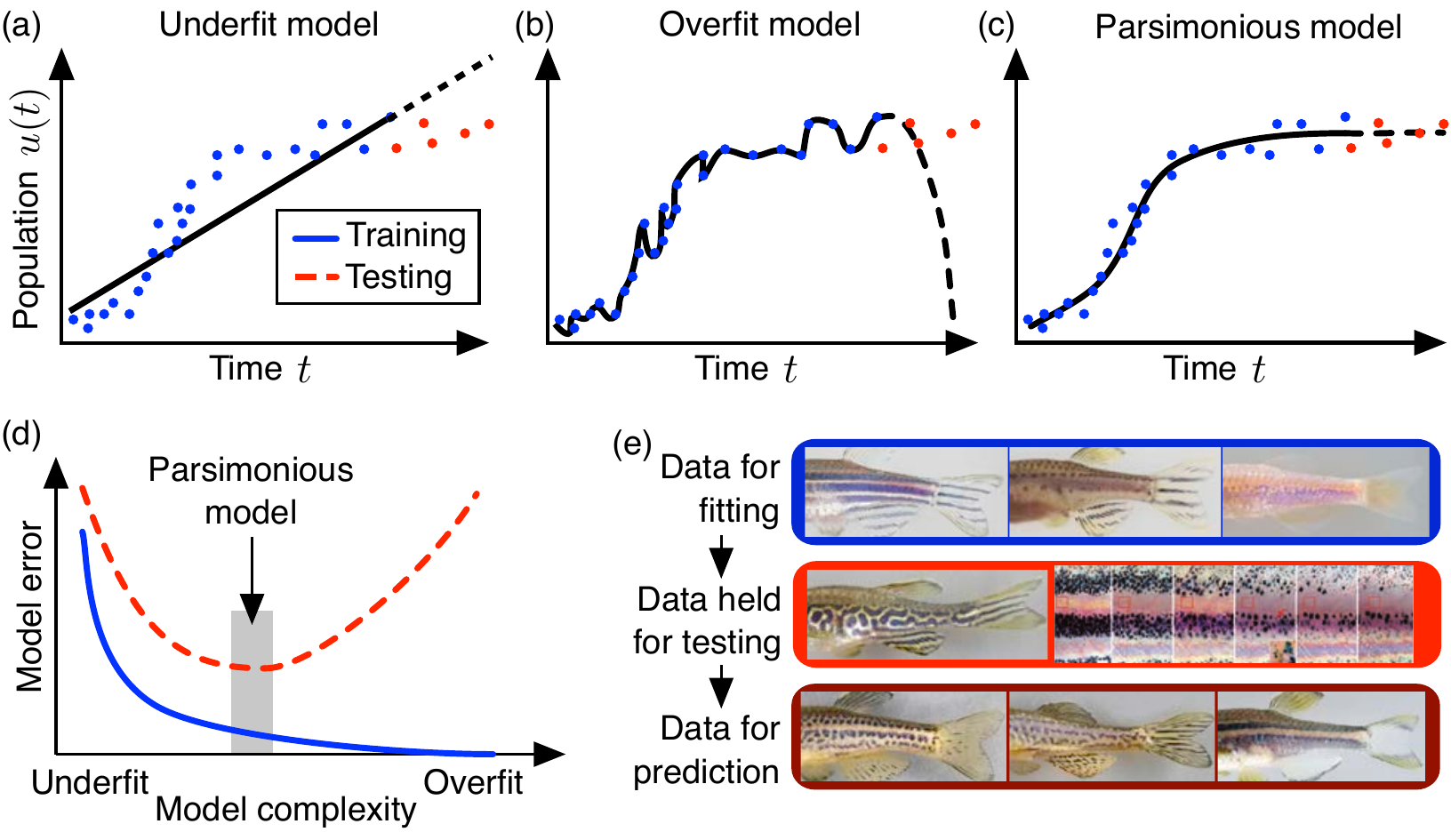}
\caption{Balancing model and data complexity. (a) Underfit models miss meaningful features in data, (b) overfit models include too many assumptions, and (c) parsimonious models balance model and data complexity. In (a)--(c), blue points denote data that we use to develop our model and fit parameters, and red points denote our testing set. (d) Underfit models agree poorly with both our training and testing data, while overfit models represent our training data well and our testing data poorly \cite{brunton2019book,Mangan2017}. Parsimonious models perform well on both sets of data. (e) Creating a plan for model training/development and testing is key to data-driven modeling. This involves breaking data into sets for training and testing, a process that depends on our complex system. For example, if our goal is to understand how cells interact to form patterns in fish skin, this could mean breaking our (qualitative) data into images of fish that are well understood (and involve setting parameters in a model to zero in a clear way), more images of fish that are well understood (and involve changing the values of nonzero parameters in a clear way), and images of poorly understood fish (and involve changing parameters in unknown ways) \cite{volkening2,volkeningThesis}. The first set is used for model development, the second for testing, and the third as a place where we can make predictions \cite{volkening2,volkeningThesis}. Image (d) is based partly on \cite{brunton2019book,Mangan2017}; first row, third row, and left image in second row of (e) adapted from \cite{SinghJan} with permission from Elsevier, Copyright (2015) Elsevier Ltd.; remaining images in second row of (e) reproduced from \cite{ParTur130} with permission of The Company of Biologists, Ltd.  \label{fig:under}}
\end{figure}

In either case, it is helpful to break our data up into sets for model training (or development) and testing. Training/development data is the data that we use to build our model, specifying parameter values as well as the form of model rules and terms as appropriate. After this, we take a step back and test whether or not our model behaves well on the data that we withheld---our testing data. 
If the model does well on the testing set, we can use our model to predict future dynamics or shed light on poorly understood dynamics. If the model does not do well on the testing set, we need to return to model development. As a guiding principle, the more parameters and assumptions that we build into a model, the more that it needs to be able to reproduce in order to have predictive value\footnote{It is worth noting that data-driven models can be used for many purposes, including describing, explaining, or predicting, and Shmueli discusses these goals from a statistical perspective in \cite{Shmueli2010}.}.

Figure~\ref{fig:under}(c)--(d) highlights two concepts that are related to balancing model and data complexity: underfitting and overfitting. For illustrative purposes, I consider the example of population growth of some organism in time, given some (imperfect) measurements of the number of organisms at discrete time points. At one extreme, I could assume a linear relationship between population size and time, fitting a line to the data. This involves few parameters, and the difference between the model and training data is high. At the other extreme, I could draw a curve that goes through every single data point \cite{SIAMmodelingComp}---this would mean introducing many parameters. In terms of these models' ability to approximate population size at some new time in our testing set, neither will do well \cite{Mangan2017}. A better model lies somewhere in between these two extremes. What we are after is a ``parsimonious" model \cite{Mangan2017,brunton2019book}: a model that it is supported by our data  and no more complex than it needs to be.

Building predictive, data-appropriate models that avoid overfitting and have strong predictive value looks different based on the problem and relies on domain expertise. (See e.g., \cite{SmithBook,brunton2019book} for a more detailed discussion of methods---I focus on broad concepts here.) If our goal is to understand social-media engagement in time, for example, we might build a gray-box model driven by some data $\{w_i\}_{i=1,....,T}$, where $w_i$ is the number of accounts on a social-media platform on day $t = i$. As one approach, we could split the data into a training set $\{w_i\}_{i=1,....,\tilde{T}}$ and a testing set $\{w_i\}_{i=\tilde{T}+1,....,T}$ with $\tilde{T} < T$. We could develop our model and specify its parameters using the training set and then run our model until $t = T$ to evaluate how well it does on the testing set. If our model does well in testing, we could use it to predict future social-media engagement.

When working with qualitative data, the process of balancing model and data complexity looks different, but it is the same at its core. In Figure~\ref{fig:under}(e), I highlight the complex biological system that most of my work is on: pattern formation in zebrafish skin \cite{volkening,volkening2}. Wild-type and mutant zebrafish feature different patterns, which form through the interactions of pigment cells \cite{SinghJan,ParTur130}. Although there are some quantitative data (e.g., cell speeds), most take the form of images of fish. To build the model \cite{volkening2}, we broke these qualitative data into three sets. The first set of images contains patterns that correspond to setting specific parameters to zero in a mathematical model (e.g., setting the birth rate of black cells to zero). The second set holds some fish patterns that are relatively well understood; in this case, we know simulating them means changing parameters in a clear way (e.g., slowing domain growth). The final set contains mutant patterns that are poorly understood, patterns that form due to cell interactions that are altered in unknown ways. The first set serves as a natural model development/training set, and once we identified a model that could reproduce these fish patterns, the next step was to step back and break it down, checking if there were any ways that we could simplify the model and still maintain consistency \cite{volkening2}. ``Minimal" model in hand, we used the second set of images for testing, asking whether or not the model could reproduce data that we did not build into it. And, finally, the tested model now serves as a predictive tool to understand the fish in the third set: at this stage, we change parameters in the model with the goal of identifying altered cell interactions that may lead to mutant patterns \cite{volkeningThesis}.

In order to further improve predictive value and avoid overfitting, there are a wealth of other approaches modelers can take. We can test how uncertainty in our parameters, boundary conditions, or initial conditions affect our results, and we can explore whether other modeling approaches lead to the same conclusions. We can set parameters in our models to zero or remove rules, checking to see if our models can be made simpler without losing agreement with the training set. We can also ask questions about whether the methods that we use to judge our models influence our results: what alternative methods for measuring agreement between model output and data can we test? Throughout this process, the goal is to critically investigate our modeling assumptions as we build a parsimonious---or minimally complex---model based in our data.

\section{Illustrative Case Studies} \label{sec:case}

In the remainder of this tutorial, I turn to two case studies of complex social systems: opinion dynamics during elections (Section~\ref{sec:elections}) and pedestrian movement in crowds (Section~\ref{sec:pedestrians}). These examples illustrate some of the types of models and data from Section~\ref{sec:perspectives} in the broader framework of the challenges and choices introduced in Section~\ref{sec:modeling}. I highlight the benefits and drawbacks of different modeling choices, with the quotations from Segel and Edelstein-Keshet \cite{KeshetSegel} at the beginning of this chapter as a guide.

\subsection{Forecasting Elections} \label{sec:elections}

Political opinion dynamics are a complex social system, and here I focus on the goal of forecasting elections in the United States. Election forecasting is highly interdisciplinary, drawing on probability, geometry, dynamical systems, topology, and statistics, as well as political science, history, economics, computer science, and sociology more broadly. It naturally involves communication and public science, and different forms of data (Section~\ref{sec:elecData}). Framed by this interdisciplinarity, I illustrate a statistical, static modeling approach to elections in Section~\ref{sec:elecModel1} and a dynamic, mathematical model in Section~\ref{sec:elecModel1}. 

Many other models and methods for incorporating data into forecasts exist beyond the scope of this survey (e.g.\ data-assimilation techniques \cite{Stuart}). Election forecasting raises questions at many different scales; for example, using a compartmental model, Restrepo \emph{et al.} \cite{Restrepo2009} investigated how polling data affect whether potential voters decide to vote, and Biondo \emph{et al}. \cite{Biondo2018} developed an agent-based model to better understand how surveys influence opinions. Election forecasting is related to the broader field of opinion dynamics \cite{castellano09,porter2016}, which includes the formation and dynamics of echo chambers (e.g., \cite{Sasahara2021,Evans2018,Cinelli2021}) and polarization (e.g., \cite{Miller2020,Yang2020}). There are many approaches to opinion formation, such as voter models \cite{PhysRevLett.112.158701,Braha} and threshold models \cite{LehmannSune2018}.

\begin{figure}[t]
\includegraphics[width=\textwidth]{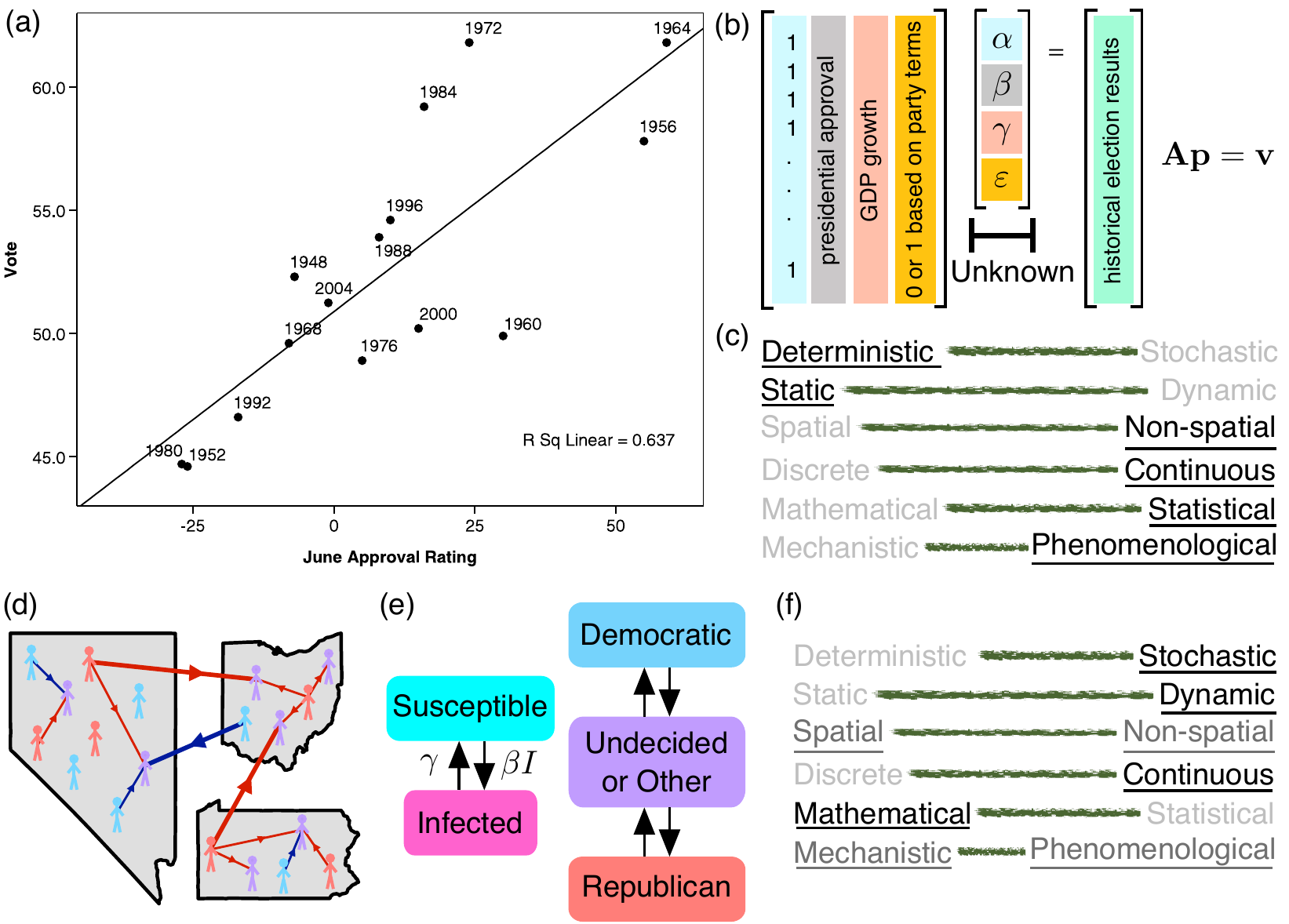}
\caption{Example approaches to forecasting U.S.\ elections. (a) The president's June approval rating and the percentage of the national vote for their party tend to be related \cite{Abramowitz2008}. (b) Abramowitz's \cite{Abramowitz2008} model is driven by fundamental data. (c) This statistical model is deterministic and continuous in the sense that, once the parameter values are set, the result is one prediction of the national vote for the incumbent party (a continuous number between $0$ and $100$). (d) In a network model, one could investigate the interactions between undecided (purple), Republican (red), and Democratic (blue) voters, as I illustrate in this cartoon. Because networks \cite{porter2016,Strogatz2001,NewmanNetworks} are the focus of another chapter in this volume, I do not cover them; I suggest the lectures by Brooks and DeFord in our Short Course \cite{BrooksVideo,DeFordVideo}. (e) Compartmental modeling \cite{HethcoteReview,Diekmann,ccc} involves grouping individuals into categories and investigating how folks change compartments. (f) The compartmental model \cite{VolkeningElec} is a stochastic mathematical approach to forecasting elections. It is spatial in the sense that it produces state-level forecasts, and non-spatial in the sense that it does not track the locations of individual voters. Image (a) reproduced from \cite{Abramowitz2008} with permission, published by Cambridge University Press, and Copyright (2008) The American Political Science Association; images (d)--(e) adapted from \cite{VolkeningElec}. \label{fig:elections}}
\end{figure}

Because elections receive attention so widely and forecasts have the potential to impact turnout, the example of election forecasting highlights a place in complex-systems research where carefully presenting the results of data-driven models is especially important. Communicating probabilistic forecasts in a tangible, interpretable way itself leads to questions, and I suggest \cite{Gelman2020,Franconeri2021} for further discussion about visualizing and communicating uncertainty. Election forecasting also presents interesting challenges when it comes to evaluating model success and forecast accuracy \cite{Gelman2020}, as I mentioned in Section~\ref{sec:data}.

\subsubsection{Election Data} \label{sec:elecData}

In terms of Step (1) in Section~\ref{sec:process}, as a starting point, the data used to build election-forecasting models include historical results, approval ratings, economic indicators, information about incumbency, and polls \cite{HUMMEL2014123,SilverBook,Abramowitz2008}. Analysts often separate these data into two types: polls and ``fundamental data" (or ``fundamentals"). Fundamental data are the data from which voters may form their opinions and determine how they will vote \cite{Gelman}; for example, economic data fall into the fundamentals category. Regardless of the type, all data come with challenges: data may not go back in time as far as a we would hope or may not be as fine-scale as we would like (e.g., data at the national or state level, rather than the district level).

For forecasts that depend on historical data, one assumption is that the past and the future will behave similarly. Modeling with fundamental data allows forecasters to produce early predictions, prior to when accurate polls may be available \cite{HUMMEL2014123,Linzer}. However, opinions are dynamic---both across years and within the same election year---and past elections may not be representative of how voters will behave in the future. On the other hand, it is not always clear whether shifts in polls in a given year represent real shifts in opinion or just differences in pollster methods \cite{Gelman, Wlezien,Jackman}. Moreover, polling data are often bias \cite{Jackman} and adjusted in proprietary ways; for example, pollsters make decisions such as how to define ``likely voters". Polling data can be spotty, with some states being polled more frequently than others \cite{Linzer}. Adding another layer of complexity, pollster herding is a phenomenon in which polling organizations adjust their results when their data do not align with other polls \cite{Herding,Clinton2013,Gelman2020}.

\subsubsection{Example Statistical Approach} \label{sec:elecModel1}

In Figure~\ref{fig:elections}(a), I reproduce a plot from \cite{Abramowitz2008} of net presidential approval ratings\footnote{This is defined as approval minus disapproval (see \cite{Abramowitz2008} for details), so it can be negative.} in June versus the percentage of the vote that went for the incumbent president's party in November of the same year. This motivates a statistical modeling approach to forecasting U.S.\ elections that is driven by fundamental, historical data. As an example of such an approach, I highlight some of the ideas in Abramowitz's ``time-for-change" model \cite{Abramowitz2008,Abramowitz1988}:
\begin{align}
 \begin{bmatrix} v_1 \\ v_2 \\ v_3 \\ . \\ . \\ v_m\end{bmatrix}
 = 
  \begin{bmatrix}
   1 & a_1 & g_1 & c_1 \\
   1 & a_2 & g_2 & c_2 \\
   1 & a_3 & g_3 & c_3 \\
   . & . & . & .\\
   . & . & . & .\\
   1 & a_m & g_m & c_m \\
   \end{bmatrix}
     \begin{bmatrix}
   \alpha \\ \beta \\ \gamma \\ \varepsilon
   \end{bmatrix}, \label{eq:abram}
\end{align}
where $v_i$ is the percentage of the national vote that went for the presidential candidate from the incumbent party in the $i$th election in the data set; $m$ is the number of years for which data is available; $a_i$ is a measurement of presidential approval before the $i$th election; $g_i$ includes information about economic growth in the year leading up to the $i$th election; and $c_i$ is a variable related to incumbency. Once the parameters $\alpha, \beta, \gamma,$ and $\varepsilon$ are determined from historical data (e.g., using regression), the time-for-change model \cite{Abramowitz2008,Abramowitz1988} can predict an election $m+1$ by computing $v_{m+1} = \alpha + \beta a_{m+1} + \gamma g_{m+1} + \varepsilon c_{m+1}$.

Equation~\eqref{eq:abram} has the general form $\textbf{v} = \textbf{A}\textbf{p}$, where $\textbf{p}$ corresponds to parameters, $\textbf{A}$ contains fundamental data, and $\textbf{v}$ holds $m$ past election outcomes. If we were to introduce more types of historical data, the number of parameters $n$ would grow. With more parameters, we would expect to get a better fit between the model predictions and past election results. As Figure~\ref{fig:under} highlights, however, this does not necessarily correspond to better predictions of future elections, since allowing $n$ to become too large can lead to overfitting. This raises questions about model complexity. How many kinds of fundamental data should a modeler include? How many terms in the model is the ``right" number of terms?

To address these questions, we need to define what a good model means and choose how to measure error. For example, consider the function \cite{brunton2019book}:
\begin{align}
E(\textbf{p}) &= \underbrace{||\textbf{A}\textbf{p} - \textbf{v}||_2}_{\text{least-squares term}} + \underbrace{\lambda_1 ||\textbf{p}||_1}_\text{LASSO term} + \underbrace{\lambda_2 ||\textbf{p}||_2}_\text{ridge-regression term}. \label{eq:error}
\end{align}
We can minimize $E(\hat{\textbf{p}})$ to find the parameter values most consistent with our data:
\begin{align*}
\textbf{p} &= \underset{\hat{\textbf{p}}}{\mathrm{argmin}}~E(\hat{\textbf{p}}).
\end{align*} 
When $\lambda_1 = \lambda_2=0$ in Equation~\eqref{eq:error}, $E(\textbf{p})$ is the least-squares difference between the model's predictions and the election outcomes under the parameters $\textbf{p}$. This method for measuring goodness-of-fit is sensitive to variability \cite{brunton2019book}. If $\lambda_1 >0$ and $\lambda_2=0$, we instead implement LASSO regression \cite{Tibshirani1996}, which selects sparse models and helps prevent overfitting by forcing some parameters to zero \cite{brunton2019book}. When $\lambda_1>0$ and $\lambda_2 >0$, Equation~\eqref{eq:error} corresponds to elastic-net regularization. 

Importantly, $\lambda_1$ and $\lambda_2$ provide a means of calibrating model complexity. We can choose to minimize Equation~\eqref{eq:error} for different values of the hyper-parameters $\lambda_1$ and $\lambda_2$, resulting in different models (in the form of the parameter values $\textbf{p}$) for each choice. Information criteria, such as Akaike information criteria (AIC) and Bayes information criteria (BIC), can come in handy to select the best model from among these alternatives \cite{Mangan2017,AIC1,AIC2,BIC}. \emph{The Economist}'s 2020 forecasts \cite{Economist}, for example, depend in part on a statistical model of the form $\textbf{A}\textbf{p} = \textbf{v}$ with a matrix $\textbf{A}$ that contains many types of fundamental data. To help prevent overfitting, \emph{The Economist} \cite{Economist} team combines leave-$1$-out cross validation \cite{brunton2019book} and elastic-net regularization with a range of $\lambda_1$ and $\lambda_2$.

Broadly, leave-$k$-out cross validation is a means of breaking data into training and validation sets. To implement this method, one removes $k$ samples of the training data; the removed data then becomes the validation set, and the remaining data is used for training \cite{brunton2019book}. For example, if $k=1$ in the presidential election setting and the available data are for the years 2004, 2008, 2012, and 2016, one first removes one year of data (e.g., the 2012 data). The next step is determining the parameter values $\textbf{p}_\text{2012}$ that result from fitting based on the data for the remaining years (2004, 2008, and 2016, in this example). Repeating this for the other years leads to four sets of parameter values. One option is to define the final parameter values $\textbf{p}$ as the mean of these four sets of parameters. Other approaches to testing and validation include $k$-fold cross validation \cite{brunton2019book}.

In Figure~\ref{fig:elections}(b)--(c), the statistical, phenomenological approach of this section has benefits and drawbacks, like all models do. Because it is driven by fundamental data, the time-for-change model \cite{Abramowitz2008,Abramowitz1988} is not dependent on noisy polling data; instead, it is able to generate forecasts as early as approval, economic, and incumbency data are available. Moreover, this model is simple and has few parameters. On the other hand, the model \cite{Abramowitz2008,Abramowitz1988} in Figure~\ref{fig:elections}(b) is static, and it does not add mechanistic understanding of what causes opinions to change in time during an election year.

\subsubsection{Example Dynamical-Systems Approach}\label{sec:elecModel2}

As a more mechanistic approach, one example is the mathematical model \cite{VolkeningElec} that my collaborators and I developed for forecasting U.S.\ elections. This model, driven by polling data \cite{HuffPostPollster,HuffPostAPI,RealClearPoliticsData,LatestPolls}, has a compartmental Susceptible--Infected--Susceptible (SIS) model at its core. Compartmental modeling is a widely used method for describing disease dynamics (e.g., \cite{Kermack700,Kermack55,Kermack94,HethcoteReview,Diekmann,ccc}), and it has also been applied to social contagions (e.g., \cite{IdeaSpread,FrenchRiots}). The central concept is that the population of interest can be grouped into compartments\footnote{This general structure is very flexible: for example, in Figure~\ref{fig:pedestrians}(d), I highlight one way that compartmental modeling could be used to describe pedestrian dynamics. Here the compartments are leading pedestrians moving to the right, following pedestrians moving to the right, leading pedestrians moving to the left, and following pedestrians moving to the left. If we are mainly interested in understanding how many leaders and followers are present, this approach could suffice. We might consider the transition of left-moving leaders to left-moving followers as dependent on interactions with other leaders.}. In the SIS setting (Figure~\ref{fig:elections}(e)), there are two compartments: susceptible and infected. Susceptible individuals become infected through interactions with infected folks (i.e., transmission), and infected individuals recover, becoming susceptible. If we track the fraction of the population that is susceptible or infected in time, the result is a gray-box model in the form of differential equations.

In the approach \cite{VolkeningElec}, we adapt the traditional SIS compartmental model by introducing two ``contagions" (Democratic and Republican voting inclinations) and replacing susceptible individual with undecided or other voters. For each state or region $i$, we track the fraction of undecided $S^i(t)$, Democratic $I^i_\text{D}(t)$, and Republican $I^i_\text{R}$ voters in time according to the stochastic ordinary differential equations:
\begin{align} 
 	{dI_\text{D}^i}(t) &= \underbrace{-\gamma^i_\text{D} I^i_\text{D}}_\text{Dem.\ recovery}dt+\underbrace{\sum_{j=1}^{M} \beta^{ij}_\text{D} \frac{N^j}{N} S^iI^j_\text{D}}_\text{Dem.\ transmission}dt + \underbrace{\sigma dW^i_\text{D}(t)}_{\text{uncertainty}}\,, \label{eq:sde1} \\
 	{dI_\text{R}^i}(t) &= \underbrace{-\gamma^i_R I^i_\text{R}}_\text{Rep.\ recovery}dt + \underbrace{\sum_{j=1}^{M} \beta^{ij}_\text{R} \frac{N^j}{N} S^iI^j_\text{R}}_\text{Rep.\ transmission}dt + \underbrace{\sigma dW^i_\text{R}(t)}_\text{uncertainty}\,, \label{eq:sde2} 
\end{align}
where we use that $S^i(t) = 1 - I^i_\text{D}(t) - I^i_\text{R}(t)$ to reduce the number of equations. Here $I_\text{D}^i, I_\text{R}^i,$ and $S^i$ are stochastic processes; $W^i_\text{D}$ and $W^i_\text{R}$ are Wiener processes; $M$ is the number of states or regions; $N^j$ is the number of voting-age individuals in state $j$; and $N$ is the total number of voting-age individuals across our $M$ regions. This model involves the simplifying assumption that we can bin voters as Democratic, Republican, or undecided. Bounded-confidence and related models (e.g., \cite{Weiss2014,Deffuant2000,Hegselmann2002,Brooks2020}) account for opinions existing on a continuous spectrum.

The parameters in Equations~\eqref{eq:sde1}--\eqref{eq:sde2} call for special attention. There are $2\times M$ parameters $\{\gamma_\text{D}^i,\gamma_\text{R}^i\}_{i = 1, ...,M}$ that describe the rates at which committed voters become undecided. There are also $2\times M^2$ parameters $\{\beta^{ij}_\text{D},\beta^{ij}_\text{R}\}_{i,j = 1,...,M}$ for the rates at which Democratic (Republican) voters in state $j$ ``infect" undecided voters. To find the values of these parameters, we \cite{VolkeningElec} relied on polling data. For the ODEs associated with Equations~\eqref{eq:sde1}--\eqref{eq:sde2} (with $\sigma=0$), we minimized the least-squares difference between our model output under parameters $\textbf{p}$:
\begin{align*}
\textbf{X}(t_k; \textbf{p}) &= [I^1_\text{R}(t_k; \textbf{p}),...,I^M_\text{R}(t_k; \textbf{p}),I^1_\text{D}(t_k; \textbf{p}),...,I^M_\text{D}(t_k; \textbf{p}),S^1(t_k; \textbf{p}),...,S^M(t_k; \textbf{p})],
\end{align*}
and the averaged state- or region-level polling data:
\begin{align*}
\textbf{x}(t_k) &= [R^1(t_k),...,R^M(t_k),D^1(t_k),...,I^M_\text{D}(t_k),S^1(t_k),...,S^M(t_k)],
\end{align*}
where $k = 1,..., T$ with $T$ months of polling data considered. The parameter values are different for each election year and race, depending on the associated polls.

The goal of forecasting elections provides a natural means of building and testing a model. By using only the polling data (but not the election results) for past races, we can test Equations~\eqref{eq:sde1}--\eqref{eq:sde2} by retroactively forecasting past elections \cite{VolkeningElec}. For the statistical model in Figure~\ref{fig:elections}(b), one of the challenges is selecting what types of fundamental data to include in the model, and this comes down to determining what parameters are zero or nonzero. In contrast, for the mathematical model here, it is more the format of the differential equations and the assumptions of an SIS-style model, rather than the values of the parameters, that we want to evaluate. Because the parameters in Equations~\eqref{eq:sde1}--\eqref{eq:sde2} depend only on the polls for a given election year, this model can be tested by applying it to forecast previous elections, one at a time. This step in some sense combines model training and validation together. In terms of predictions, there is also a natural---and high-stakes---opportunity: the model can be used fo forecast upcoming elections.

One of the benefits of the continuous, stochastic mathematical model in Equations~\eqref{eq:sde1}--\eqref{eq:sde2} is that it includes some mechanistic hypotheses about opinion dynamics. The model \cite{VolkeningElec} is also dynamic in time; see Figure~\ref{fig:elections}(f). Once polls becomes available, Equations~\eqref{eq:sde1}--\eqref{eq:sde2} can forecast a new U.S.\ election with parameters that are specific to that election. However, opinion dynamics are not the same as biological disease transmission. Instead, we might think of the transmission terms in Equations~\eqref{eq:sde1}--\eqref{eq:sde2} as capturing interactions between committed voters in state $j$ and undecided voters in state $i$ in a phenomenological way. These interactions could be direct (e.g., via conversations between a committed voter in one state and an undecided voter in another state) or indirect (e.g., through news coverage). As another drawback, the model \cite{VolkeningElec} has many more parameters than the statistical approach \cite{Abramowitz1988,Abramowitz2008}.

\subsection{Modeling Pedestrian Movement} \label{sec:pedestrians}

\begin{figure}[t]
\includegraphics[width=\textwidth]{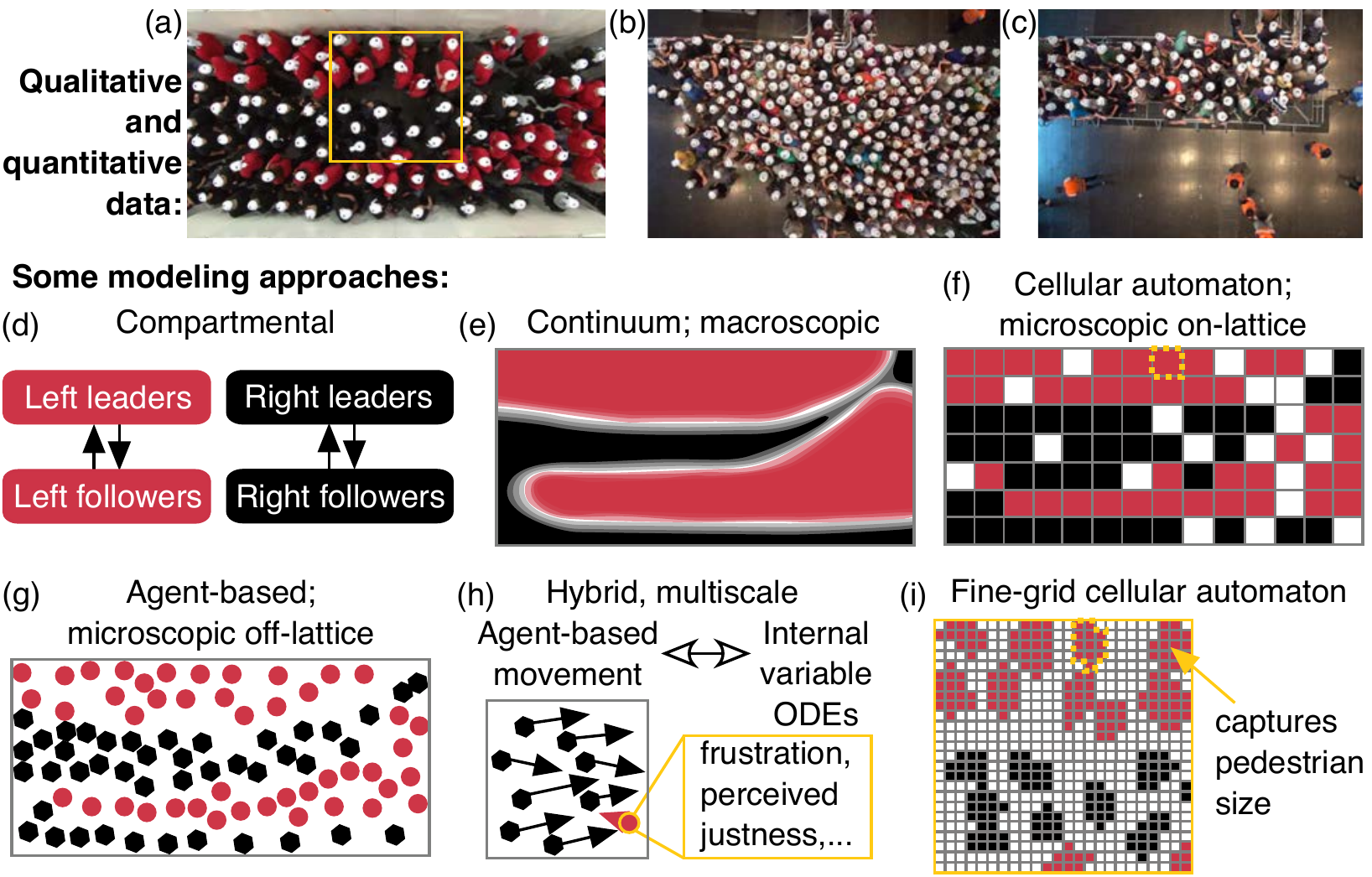}
\caption{Example modeling approaches to pedestrian movement. Experiments in settings such as (a) bidirectional movement in corridors \cite{Zhang2012}, (b) movement through through an entrance without broader spatial barriers \cite{Sieben2017}, and (c) movement through an entrance with spatial constraints \cite{Sieben2017} produce both quantitative and qualitative data. There are many approaches that we could take to describe lane formation in (a).  In (a) and (d)--(i), red denotes pedestrians moving to the left and black denotes study participants moving to the right. (d) Compartmental models, discussed in Section~\ref{sec:elecModel2} could, for example, track the fraction of study participants who are following others or leading lines; this approach is non-spatial. (e) Macroscopic models track pedestrian density and generally take the form of PDEs. (f) Microscopic on-lattice models consider the positions of individuals in discrete space and involve stochastic, computational rules. (g) Microscopic on-lattice models track the positions of individuals in continuous space through coupled differential equations. (h) Hybrid, multiscale approaches come in many forms; for example, we could couple an agent-based model of pedestrian movement with a compartmental model describing the feelings within each pedestrian, which, in turn, influence their movement. (i) Fine-grid cellular automaton models use multiple grid squares to represent each pedestrian, providing a more detailed perspective on pedestrian position. Images (a)--(c) adapted (cropped) from \cite{Sieben2017} and licensed under CC-BY 4.0 (\url{https://creativecommons.org/licenses/by/4.0/}). \label{fig:pedestrians}}
\end{figure}

Crowds of people exhibit rich collective behavior, including lane formation and oscillating flows \cite{Helbing1995,helbing2005self,Helbing2009,Sieben2017}. For example, as I show in Figure~\ref{fig:pedestrians}(a), pedestrians may form lanes when two groups walk in opposing directions in a narrow corridor. Like the application of election forecasting in Section~\ref{sec:elections}, studying the dynamics of crowds touches on many fields, including engineering, sociology, psychology, physics, computer science, and mathematics \cite{Bellomo2016,Sieben2017,Bode2019,Helbing2009}. This interdisciplinarity stems from the goals that can motivate models of pedestrian movement. Researchers may be interested in designing functional buildings, testing how guidelines influence disease transmission in a crowd, developing methods to improve evacuation in emergency settings, or something else. Here I focus on the goal of understanding under what conditions lanes emerge from pedestrian interactions, and I assume accounting for the spatial organization of individuals in time is important.

For this tutorial, I use crowd movement as a venue for discussing approaches to modeling agent behavior in space, and highlighting some challenges associated with qualitative data (Section~\ref{sec:pedestriansData}). Pedestrian movement provides an opportunity to illustrate a range of gray-box, spatial models, including continuum models, cellular-automaton perspectives, and agent-based approaches (Section~\ref{sec:pedestriansModel1}). There are other data-driven approaches to crowd dynamics, and I highlight \cite{Bode2019} for a review of statistical models. From the perspective of building simplified models (in particular, models that do not include concepts from social psychology \cite{Sieben2017}), similar challenges and approaches arise in diverse examples of pattern formation and self-organization, including migrating cells (e.g, \cite{Buttenschon2020,VolkeningRev,Giniunaite2020,Merks2017,Glazier1993}), animal aggregations (e.g., \cite{Couzin2002,Parrish1999,Lukeman2010keshet}), swarming locusts  (e.g., \cite{Ariel2015,Bernoff2020,Bernoff2011}), and more general agents interacting in space (e.g., \cite{Chuang2007,Vicsek1995,Levine2001,DOrsogna2006,Mogilner1999,Topaz2006,Carrillo2016}).

\subsubsection{Pedestrian Data} \label{sec:pedestriansData}

Data on pedestrian movement comes in quantitative and qualitative forms, including measurements of velocity \cite{Zhang2012}, questionnaires about pedestrian experience \cite{Sieben2017}, and images of crowds \cite{Bosse2011}. This information may stem from observations in the field or in controlled lab settings. For example, Zhang \emph{et al.} \cite{Zhang2012} performed a series of experiments in which study participants were instructed to move through corridors of different widths. As I show in Figure~\ref{fig:pedestrians}(a), participants in red were asked to move to the left through the corridor, and pedestrians wearing black were asked to move to the right. Lanes---visible as red and black stripes in Figure~\ref{fig:pedestrians}(a)---emerged from the interactions of the pedestrians in some settings \cite{Zhang2012}. In addition to this qualitative data, the experiments \cite{Zhang2012} produced trajectories of each participant's position, along with measurements of velocity and density.

As another example, Sieben \emph{et al.} \cite{Sieben2017} performed a series of experiments to better understand how pedestrians respond to different barriers as they seek to pass through an entrance. The setups \cite{Sieben2017} in Figure~\ref{fig:pedestrians}(b)--(c) are meant to represent what might happen when people are entering a concert venue. After extracting the positions of the white caps worn by pedestrians, Sieben \emph{et al}. \cite{Sieben2017} collected trajectories of individuals. The authors \cite{Sieben2017} also asked the study participants questions about their experience of walking through the entrance before and after watching a video of the experiment. This survey \cite{Sieben2017} produced data on how comfortable the heterogeneous participants reported feeling and how just they felt the entrance process was, among other things.

When we view Figure~\ref{fig:pedestrians}(a), the presence of stripes is striking; while it is not as visible in Figure~\ref{fig:pedestrians}(c), the trajectories of pedestrian movement that Sieben \emph{et al}. \cite{Sieben2017} extracted from these experiments also show lanes in some cases. This highlights one of the challenges associated with spatial complex systems: many of the features in Figure~\ref{fig:pedestrians}(a)--(c) are qualitative. We may see stripes, but how do we define these stripes objectively and quantitatively in large sets of images? At different timepoints in the experiment (see the videos in the supplementary material of \cite{Sieben2017}), the stripes are not as clear and do not extend across the full length of the corridor. How do we define stripe width or the time when bands start or end along the length of the corridor? The qualitative nature of data in spatial complex systems presents new challenges when fitting and testing models.

\subsubsection{Example Spatial Modeling Approaches}\label{sec:pedestriansModel1}

Figure~\ref{fig:pedestrians} shows some approaches to spatial modeling of complex systems, including crowd movement, at different levels of detail. Here I focus on introducing some broad gray-box, mathematical modeling approaches that we could take to study lane formation, rather than discussing specific references. (See the reviews \cite{Bellomo2016,Schadschneider2018,Duives2013} and references therein for more information about crowd dynamics.) These approaches---namely macroscopic, microscopic on-lattice, microscopic off-lattice, and hybrid (e.g., \cite{Kneidl2013}) models---are used to study a wealth of spatial dynamics. There are also many perspectives that I do not discuss, including mesoscopic (e.g., \cite{festa2018kinetic,bellomo2013microscale}) and game-theoretic (e.g., \cite{Dogbe2010,lachapelle2011mean,Bailo2018}) approaches.

Macroscopic, continuum models of pedestrian movement often take the form of partial differential equations (PDEs). As I show in Figure~\ref{fig:pedestrians}(e), this approach stems from a zoomed-out perspective: instead of tracking the locations of individuals in Figure~\ref{fig:pedestrians}(a), continuum models describe the evolution of density in time. If we make the assumption that there are two populations in our corridor example, a continuum model would track the density $r(\textbf{x},t)$ of ``red-shirt-wearing" and $b(\textbf{x},t)$ ``black-shirt-wearing" pedestrians in space $\textbf{x}$ and time $t$. One benefit of macroscopic models is that they are often analytically tractable, and they provide a broad perspective on overarching features that may be at work in a complex system. These models often have few parameters, and researchers can perform bifurcation analysis to understand how these parameters influence group dynamics. The drawback is that PDE approaches may simplify the complex dynamics of heterogeneous pedestrians significantly, and it can be challenging to relate the few parameters in these models to specific agent behaviors.

In contrast to macroscopic models, microscopic approaches focus on the positions or features of individuals, and two prominent frameworks are on-lattice and off-lattice models. These models provide more detailed perspectives at the scale of individual agents, which comes at the cost of more parameters. Spatial modeling is a place where vocabulary differs some between fields, particularly in the case of microscopic models. Depending on one's perspective, the microscopic models in Figure~\ref{fig:pedestrians}(f)--(e) may be described as individual- or agent-based models (IBMs or ABMs), since these models track changes in the positions of agents. The term ``agent-based" also refers to more detailed models such as \cite{Bosse2011,Bosse2009,Tsai2011}. Miller and Page \cite{Miller2007} describe agent-based models as ``bottom-up" approaches, because the starting point is interactions of individuals. In interdisciplinary---or even within-discipline---conversations, I suggest asking questions to clarify what folks mean by ABMs and IBMs in their setting.

Microscopic on-lattice (cellular automaton) models consider space as a lattice, and pedestrians can either occupy or not occupy positions on a grid (e.g., \cite{Burstedde2001,Varas2007,Blue2001}); see Figure~\ref{fig:pedestrians}(f). Movement, as well as arrival and exit, takes the form of stochastic, computational rules. Notationally, we could denote whether the grid square in row $i$ and column $j$ at time $t_k$ is red (i.e., containing a pedestrian moving to the left in Figure~\ref{fig:pedestrians}(a)), black (i.e., holding a right-moving pedestrian), or white (empty) by:
\begin{align*} 
x_{i,j}(t_k) &= \begin{cases}
-1 & \text{~if  grid square is red} \\
0 & \text{~if grid square is empty} \\
1 & \text{~if grid square is black}.
\end{cases}
\end{align*}
For example, to model right-traveling pedestrians stepping to the side to avoid collisions with left-moving study participants, we might select a grid square $(i,j)$ uniformly at random from Figure~\ref{fig:pedestrians}(a) and implement the rule:
\begin{align}
&\text{if} \underbrace{~\text{$x_{i,j}(t_k) = -1$ and $x_{i,j+1}(t_k)  = 1$}}_\text{conditions for a head-on collision} \text{~and~} \underbrace{\text{$x_{i+1,j}(t_k)  = 0$}}_\text{space available}, \nonumber \\
&\text{then~}\underbrace{\text{$x_{i,j}(t_{k+1})  = 0$ and $x_{i+1,j}(t_{k+1}) = 1$ with probability $p$}}_\text{pedestrian may step to the side}.\label{eq:ped}
\end{align}
In one time step, we could iterate through a random perturbation of all of the grid squares, implementing this and other model rules. There are many choices and parameters in Rule~\eqref{eq:ped}, including the choice of probability $p$ and the choice of neighborhoods considered (e.g., why should the pedestrian at space $(i,j)$ only look one grid step ahead to space $(i,j+1)$? Maybe $(i,j+2)$ is more appropriate?). 

Microscopic off-lattice models (e.g., \cite{Helbing1995,helbing2005self}), in comparison, assume that individuals move continuously in space; see Figure~\ref{fig:pedestrians}(g). In this case, movement is modeled through coupled ordinary or stochastic differential equations, for example, of the form:
\begin{align} 
\frac{d\textbf{V}_i}{dt} &= \underbrace{\textbf{g}(\textbf{X}_i, \textbf{V}_i)}_\text{pedestrian $i$'s inherent goals} + \underbrace{\sum_{j=1}^N \textbf{f}(\textbf{X}_i, \textbf{X}_j, \textbf{V}_i, \textbf{V}_j)}_\text{interactions between pedestrians} \label{eq:ped2}\\
\frac{d\textbf{X}_i}{dt} &= \textbf{V}_i,\label{eq:ped3}
\end{align}
where $\textbf{X}_i(t)$ is the position of the $i$th pedestrian (e.g., a point mass marking the $(x,y)$ coordinates of the pedestrian's center of mass) and $\textbf{V}_i(t)$ is that pedestrian's velocity. So called ``social-force" models are a prominent off-lattice microscopic approach to pedestrian dynamics \cite{helbing2005self,Helbing1995}. In both on-lattice and off-lattice models, arrival and exit of pedestrians from either side of the corridor in Figure~\ref{fig:pedestrians}(a) could take the form of stochastic rules. Computationally, we might assume that a new pedestrian enters the corridor at a randomly selected $(x,y)$ position near the left or right edge of Figure~\ref{fig:pedestrians}(a) with probability $\alpha \Delta t$, where $\Delta t$ is the time step of our simulations. 

While microscopic models offer detailed perspectives on the behavior of individuals and can make experimentally testable predictions, they have many more parameters than macroscopic models do. In order to avoid overfitting and improve predictive value, it is thus important to break our data into separate sets for model development and testing. For example, we could fit the parameters in the functions in Equations~\eqref{eq:ped2}--\eqref{eq:ped3} based on measurements of pedestrian--pedestrian distances and pedestrian velocities. We could specify the rates at which pedestrians enter the corridor based on empirical data, and we could use lane width to determine any unmeasurable parameters or guide the form of model rules. To test our model, we could set aside certain experiments (e.g., experiments with wider corridors) to simulate with our final model. We could, for example, use our validated model to predict how the dynamics will change when a pushier agent is introduced or when the structure of the barriers and walls in Figure~\ref{fig:pedestrians}(a)--(c) is changed.

Adding further difficulty, microscopic models are often stochastic and not analytically tractable, and they face some of the same challenges as qualitative data: how do we define and quantitatively describe the stripes in Figure~\ref{fig:pedestrians}(f)--(g) in an automated, objective way? To help address this challenge, topological techniques, especially persistent homology \cite{Otter2017mason,Edelsbrunner2008,Carlsson2020}, have recently been combined with modeling to study complex systems, including aggregation \cite{Ulmer2019topaz,Topaz2015}. Additional examples of topological data analysis applied to biological and social complex systems include \cite{Bhaskar2019,Bonilla2020,McGuirl2020,Ciocanel2021,Munch2020shape,Nardini2021arxiv} and  \cite{Feng2021tda,HickokPreprint}, respectively. Pair correlation functions \cite{Dini2018,Johnston2019pre,Treloar2014spatial} are another method for quantifying spatial data.

Depending on our goals and what our data suggest, building a hybrid, multiscale model that accounts for dynamics within pedestrians may be appropriate; see Figure~\ref{fig:pedestrians}(h). For example, in the off-lattice microscopic setting, we could introduce a variable $P_i(t)$ that tracks how frustrated each individual is based on their perceived justness of the crowd dynamics around them. We could define $P_i(t)$ by comparing the distance that pedestrian $i$ has moved toward their goal in some time interval to the estimated distance that the individuals in a local neighborhood around $i$ are moving. There are many other ways that we could define $P_i(t)$, and we could include feedback between $P_i(t)$ and how pushy pedestrians choose to be, influencing our ODEs for movement in an associated agent-based model.

As a last example, similar to cellular Potts models in biology \cite{Glazier1993,Merks2017}, fine-grid cellular automaton represent each individual with a collection of grid squares (e.g., \cite{Sarmady2010}); see Figure~\ref{fig:pedestrians}(i). These detailed approaches are appropriate when folks are interested in the spatial extent of agents. Representing each pedestrian with $N > 1$ grid squares, instead of just one as in Figure~\ref{fig:pedestrians}(f), increases the number of parameters and the time that it will take to simulate the model. This means fine-grid cellular automaton may make more sense when the goal is to describe the behavior of a few pedestrians in a detailed way; as we consider a larger crowd, agent-based or cellular automaton models become more appropriate; and, as we zoom out further into very large, densely packed crowds, macroscopic models are especially helpful. In crowd dynamics, as for other complex social systems, there are many useful modeling approaches that we could take, and it is a matter of choosing one that is parsimonious and appropriate for our goals.

\section{Conclusions} \label{sec:conclusions}

I conclude with the best piece of advice that I have been given as a modeler: don't be afraid to be wrong. In particular, developing a model that correctly describes all of the unknown, intricate details of a complex social system would come down to sheer luck, since the space of possible models is huge. This can be discouraging. Instead, I have found it freeing to recognize that all of my models have been and will continue to be ``wrong" in some sense. What matters is getting it ``wrong" in a meaningful way. By building a parsimonious model, balancing our assumptions with the amount of data available, and designing a clear method for testing the model, we can make a meaningful contribution and generate new insights despite being inevitably ``wrong" (or ``right" in a simplified way). If the first model of a complex system does not cross disciplinary boundaries, it can lay the groundwork for a bridge that brings disciplines together in the future.

Whether our starting point is a rich data set or a nearly blank space, modeling complex systems is an iterative, creative, and interdisciplinary process. It involves being aware of the choices that we are making to simplify the problem, choosing model complexity based on our data, carefully considering the bias in the data and model, and identifying a plan for model building and validation. Through data collection, model development, prediction, communication, and generating new questions, we can push the field forward, help address societal challenges, develop mathematical approaches, and bring disciplines together in new ways.

\bibliographystyle{amsalpha}
\bibliography{PSAPM_Volkening}

\end{document}